\documentclass[leqno,12pt]{article}

\usepackage{amsmath}
\usepackage{amssymb}
\usepackage{appendix}
\usepackage{amsthm}
\usepackage{graphicx}
\usepackage{setspace}

\usepackage{enumitem}
\usepackage{hyperref}

\usepackage[utf8]{inputenc}
\usepackage[english]{babel}
\usepackage{setspace}
\usepackage[margin=1in]{geometry}

\theoremstyle{definition}

\numberwithin{equation}{section}

\newtheorem{theorem}{Theorem}[section]
\newtheorem{corollary}[theorem]{Corollary}
\newtheorem{definition}{Definition}[section]
\newtheorem{assume}{Assumption}
\newtheorem{proposition}{Propositioin}[section]

\newtheorem{example}{Example}[section]

\begin{document}

\author{Hyungbin Park \thanks{hyungbin@cims.nyu.edu, hyungbin2015@gmail.com}\\
Courant Institute of Mathematical Sciences,\\ New York University\\ \\
\small{Begins at : Nov 25, 2014}\\
\small{First version: Apr 01, 2015}}
\title{The Martin Integral Representation of Markovian Pricing Kernels}

% for double-spacing footnots.
% \thanks{\protect\doublespacing The author blahblah...

\date{}   
\maketitle

\begin{abstract}
The purpose of this article is to describe all possible beliefs of market participants on objective measures under Markovian environments when a risk-neutral measure is given.
To achieve this, we employ the Martin integral representation of Markovian pricing kernels. Then, we offer economic and financial implications of this representation. This representation is useful to analyze the long-term behavior of the state variable in the market. The Ross recovery theorem and the long-term behavior of cash flows are discussed as applications.  
\end{abstract}

\section{Introduction}

Quantitative finance theory involves two related probability models: a risk-neutral measure and an objective measure. 
The risk-neutral measure determines market prices of contingent claims.
The price of the claim is the expected discounted cash flows, with expectations in the risk neutral measure.
It is distinct from the objective measure, which describes the actual stochastic
dynamics of markets, or at least the participants' beliefs about them.

A pricing kernel (or numeraire) on the objective measure is determined by the relationship between the risk-neutral measure and the objective measure.
We denote the reciprocal of the pricing kernel by $L_t.$  
In this article, a special form will be assumed in the pricing kernel known as the Markovian structure;
$$L_{t}=e^{\beta t}\,\phi(X_{t})\,\phi^{-1}(X_0)\; $$
for some positive function $\phi,$ a real number $\beta$ and a Markovian diffusion state variable $X_t.$
The function $\phi$ is called a {\em principal function} of the market.
This Markovian structure on the pricing kernel is widely accepted in many studies on asset pricing theory.
For example, in the Consumption-Based Capital Asset Pricing Model (CCAPM) in \cite{Breeden79}, \cite{Karatzas98}, the pricing kernel has the Markovian structure.

The purpose of this article is to describe all possible beliefs of market participants on objective measures under Markovian environments when a risk-neutral measure is given.
To achieve this, we employ the Martin integral representation of Markovian pricing kernels. Then, we offer economic and financial implications of this representation. This representation is useful to analyze the long-term behavior of the state variable in the market. The Ross recovery theorem and the long-term behavior of cash flows are discussed as applications.

We will show that principal functions can be characterized by the following two properties.
First, $\beta$ and $\phi$ satisfy the following second-order partial differential equation
equation
\begin{equation} \label{eqn:DE}
\frac{1}{2}\sum_{i,j=1}^{N}a_{ij}(x)\partial_{ij}\phi(x)+\sum_{i=1}^{N}k_i(x)\partial_i\phi(x)-r(x)\phi(x)=-\beta \phi(x)\,,
\end{equation}
where $X_t$ and $r(\cdot)$ satisfy Assumptions 1,2 and 3 in Section \ref{sec:Markovian pricing kernels}.
Second, 
 $$e^{\beta t-\int_{0}^{t} r(X_{s})ds}\,\phi(X_{t})\,\phi^{-1}(X_0)$$
is a martingale under the risk-neutral measure.
Any function $\phi$ satisfying two properties will be called an {\em admissible function} and can serve as a principal function.
Thus, a problem of describing the objective measures is transformed into a problem of finding admissible functions.

The Martin integral representation theory is useful to describe an admissible function.
Any admissible function $\phi$ can be expressed by
$$\phi(x)=\int_{\Gamma}k(x;y)\,d\mu_{\phi}(y)$$
where $k$ is the Martin kernel, $\Gamma$ is the admissible Martin minimal boundary and $\mu_{\phi}$ is the corresponding finite measure on $\Gamma.$
Conversely, for any finite measure $\mu$ supported on $\Gamma,$
$$\phi(x)\equiv\int_{\Gamma}k(x,y)\,d\mu(y)$$
is an admissible function.
In conclusion, there is a one-to-one correspondence between the objective measures and finite measures on $\Gamma.$
Every objective measure can be identified by the corresponding finite measure on $\Gamma.$

This representation is useful to analyze the long-term behavior of the state variable $X_t$ under the objective measure.
The `limiting distribution' of $X_t$ when $t$ goes to infinity
under the objective measure can be described by the corresponding measure $\mu.$
If a process is positive recurrent, the meaning of the limiting distribution is clear.
However, if $X_t$ is transient, then the limiting distribution does not exist in the usual sense.
While investigating the long-term behavior, this phenomenon makes transient processes harder to study.
The Martin representation is useful to overcome this.
Indeed, the measure $\mu$ can be viewed as an `extended meaning' of a limiting distribution.  
We will see this topic in Section \ref{sec:lim_disri}.

As an application, the Ross recovery theorem is discussed in Section \ref{sec:Ross_recovery}.
Recently, many authors have investigated Markovian pricing kernels in recovery literature.
Ross \cite{Ross13} argue that it is possible to determine uniquely an objective
measure from a risk-neutral measure under the Markovian structure. He assumes that there 
is a state variable $X_{t}$ with a finite number of states on 
discrete time $t\in\mathbb{N}.$ 
The author uses a combination of economic arguments and mathematical analysis.
A theory that determines an objective measure under Markovian environments from a risk-neutral measure is referred to as the {\em Ross recovery theorem}.

Many authors extended the original the Ross recovery theorem to a continuous-time setting with a time-homogeneous Markov diffusion process $X_{t}$ with state 
space $\mathbb{R}$ or $\mathbb{R}^N.$
They showed that $(\beta,\phi)$
satisfies the above second-order differential
equation \eqref{eqn:DE}.
Thus, the recovery theorem is transformed into a problem of finding a particular solution 
pair $(\beta,\phi)$ of the particular differential equation with $\phi(\cdot)>0.$ If such a solution 
pair were unique, then we could successfully recover the objective measure. Unfortunately, this 
approach categorically fails to achieve recovery because such a solution pair is never unique.
We discuss this issue in Section \ref{sec:recur_trans}.

Many authors have extended the Ross model to a continuous-time setting and have
confronted the non-uniqueness problem. To overcome the non-uniqueness problem, 
they assumed more conditions onto their models so that the differential equation \eqref{eqn:DE} has a unique solution pair satisfying the conditions.

Carr and Yu \cite{Carr12} assumed that the process $X_{t}$ is a one-dimensional time-homogeneous Markov diffusion on a {\em bounded} interval with regular boundaries at both endpoints. 
Dubynskiy and Goldstein \cite{Dubynskiy13}
explored Markov diffusion models with reflecting boundary conditions.
Walden \cite{Walden13} extended the results of Carr and Yu to the case that $X_{t}$ is an 
{\em unbounded} process. Walden proved that recovery is possible if the process $X_{t}$ is 
{\em recurrent} under the objective measure. 
Qin and Linetsky \cite{Qin14b} proved that recovery is possible 
for a Mokovian Borel right process $X_t$ if $X_{t}$ is recurrent under the objective measure.
Borovicka, Hansen and Scheinkman \cite{Borovicka14} showed that the recovery is possible if the 
process $X_{t}$ is {\em stochastically stable} under the objective measure.

The papers of Borovicka, Hansen and Scheinkman \cite{Borovicka14}, Qin and Linetsky
\cite{Qin14b} and Walden \cite{Walden13} assumed a common condition on $X_{t}$. Specifically, $X_{t}$ is {\em recurrent} under the objective measure. The mathematical rationale for 
this condition is to overcome the non-uniqueness problem of the differential equation \eqref{eqn:DE}. Indeed, 
if existent, there is a unique solution pair $(\beta,\phi)$ of the equation \eqref{eqn:DE} satisfying this condition. We will review this in Section \ref{sec:Ross_recovery}.

There are a few number of studies on Ross recovery for a transient process $X_t.$
Park \cite{Park14b} proved that recovery is possible for a one-dimensional Markovian diffusion $X_t$
if $\beta$ is known and if $X_{t}$ is {\em non-attracted} to the left (or right) boundary under the objective measure. The author also offered a graphical understanding of the Ross recovery theorem.

In this article, as an application of the Martin integral representation, we will investigate
Ross recovery with a multi-dimensional Makovian diffusion $X_t$ that is transient under the objective measure.
We first discuss why transient recovery is inevitable when studying the multi-dimensional process $X_t.$ Then, we will prove that
Ross recovery is equivalent to choosing a number $\beta$ and a finite measure $\mu$ on $\Gamma_\beta$ where $\Gamma_\beta$ is the corresponding admissible Marin minimal boundary.

For another application, we investigate long-term behavior of cash flows in Section \ref{sec:long-term}. Borovicka, Hansen and Scheinkman investigated the long-term decay (or growth) rate of cash flows in \cite{Borovicka14}. In this article, we investigate their work in the view of the Martin representation.

The following provides an overview of this article.
We first set up the Markovian economy in Section \ref{sec:Markovian pricing kernels} and then study the relationship between a risk-neutral measure and an objective measure in Section \ref{sec:transformed}.
In Section \ref{sec:recur_trans}, the notion of recurrence/transience is explored. 
In Section \ref{sec:Martin_repn}, we investigate the Martin integral representation for Markovian pricing kernels.
The long-term behavior of the state variable is demonstrated in Section \ref{sec:lim_disri}.
In Section \ref{sec:Ross_recovery}, as an application of the Martin theory, Ross recovery is discussed.
In addition, we explore the long-term behavior of cash flows in Section \ref{sec:long-term}.
The last section summarized this article.

\section{Markovian pricing kernels}
\label{sec:Markovian pricing kernels}

A {\em risk-neutral financial market} is defined as a probability space
$(\Omega,\mathcal{F},\mathbb{Q})$ having a $N$-dimensional standard Brownian motion $W_{t}:=(W_1(t),W_2(t),\cdots,W_N(t))$
with the filtration $\mathcal{F}=(\mathcal{F}_{t})_{t=0}^{\infty}$ generated by $W_{t}$.  
All the processes
in this article are assumed to be adapted to the filtration $\mathcal{F}.$
$\mathbb{Q}$ is called the risk-neutral measure of this market.
We assume that there are a state variable $X_t$ and a short interest rate process $r_t$ in the market.

Let $\mathbb{P}$ be any equivalent measure on the market with respect to the risk-neutral measure $\mathbb{Q}.$ 
In this article, the measure $\mathbb{P}$ plays the role of an objective measure. 
A special relationship between $\mathbb{Q}$ and $\mathbb{P}$ will be imposed in Assumption \ref{assume:RN}. 
Set the Radon-Nikodym derivative by 
$\Sigma_{t}:=\left.\frac{d \mathbb{Q}}{d \mathbb{P}}
\right|_{\mathcal{F}_{t}}.$
It is known that $1/\Sigma_t$ is a martingale under $\mathbb{Q}.$ 
Using the martingale representation theorem, we can write in the SDE form 
$$d\left(\frac{1}{\Sigma_{t}}\right)=\frac{1}{\Sigma_{t}}\,\rho_t\cdot dW_{t}$$ for some $\rho_{t}$.
 It is well-known that $B_{t}$ defined by
\begin{equation}\label{eqn:Girsanov}
dB_{t}:=-\rho_{t}\cdot dt+dW_{t}
\end{equation}
 is a $N$-dimensional Brownian motion under $\mathbb{P}.$

\begin{assume} \label{assume:interest_rate}
The short interest rate $r_t$ is determined by $X_{t}.$ More precisely, 
there is a continuous positive function
$r(\cdot)$ such that $r_{t}=r(X_{t}).$
\end{assume}
\noindent Define {\em the reciprocal of the pricing kernel} by $L_{t}=e^{\int_0^tr(X_s)\,ds}/\Sigma_{t}.$  Then 
\begin{equation} \label{eqn:RN_SDE}
\begin{aligned}
\frac{dL_{t}}{L_t}&=(r(X_{t})+|\rho_{t}|^2)\,dt+\rho_{t}\cdot dB_{t}\\
&=r(X_{t})\,dt+\rho_t\cdot dW_{t} 
\end{aligned}
\end{equation}
is obtained.

We define a pricing operator $P_t$ by
\begin{equation}\label{eqn:pricing_oper}
P_t(f):=\mathbb{E}^{\mathbb{Q}}\left[e^{-\int_0^tr(X_s)\,ds}f(X_t)\right]\;.
\end{equation}

\begin{assume} \label{assume:X}
The state variable $X_t=(X_1(t),X_2(t),\cdots,X_N(t))$  is a $N$-dimensional
time-homogeneous Markov diffusion process satisfying the following stochastic differential equation.
$$dX_i(t)=k_i(X_t)\,dt+\sigma_i(X_t)\cdot dW_t\quad\text{ for }\; i=1,2,\cdots,N,$$ 
where $\sigma_i(\cdot)=(\sigma_{i1}(\cdot),\sigma_{i2}(\cdot),\cdots,\sigma_{iN}(\cdot)).$
Let $a_{ij}:=\sum_{k}\sigma_{ik}\sigma_{kj}.$ 
We assume that the pricing operator $P_t$
induces the infinitesimal generator $\mathcal{L}$ on a domain $D\subseteq\mathbb{R}^N$
$$\mathcal{L}h(x)=\frac{1}{2}\sum_{i,j=1}^{N}a_{ij}(x)\partial_{ij}h(x)+\sum_{i=1}^{N}k_i(x)\partial_ih(x)-r(x)h(x)$$
which satisfies that 
$a_{ij}$ and $k_i$ are continuous for all $i,j$ on $D$ and that 
$\sum_{i,j}a_{ij}(x)v_iv_j>0$ for all $x\in D$ and $v\in\mathbb{R}^N-\{0\}.$
\end{assume}

\begin{assume}\label{assume:RN} \textnormal{(Markovian Pricing Kernel)}\newline
Assume that the relationship between $\mathbb{Q}$ and $\mathbb{P}$ is determined by a positive
a positive function $\phi\in C^{2}(\mathbb{R}^{N})$ and a number $\beta.$
More precisely,
$L_t$ is expressed by
\begin{equation} \label{eqn:RN_X}
L_{t}=e^{\beta t}\,\phi(X_{t})\,\phi^{-1}(X_0)\; .
\end{equation}
\end{assume} 
\noindent Equivalently, $\Sigma_t$ is written by
$$e^{-\beta t+\int_{0}^{t} r(X_{s})ds}\,\phi^{-1}(X_{t})\,\phi(X_0)\;.$$
In this case, we say that $\beta, \phi$ are the {\em principal factor} and the {\em principal function} with respect to the measure $\mathbb{P},$ respectively.
The pair $(\beta,\phi)$ is called the {\em principal pair} corresponding to the measure $\mathbb{P}.$
As an example, the standard argument of the CCAPM says that
the pricing kernel is expressed by
$$L_t=e^{\beta t}\frac{U'(c_0)}{U'(c_t)}$$
where $U$ is the utility of the representative agent, $\beta$ is the discount factor and $c_t$ is the aggregate consumption process.

\section{Transformed measures}
\label{sec:transformed}
In this section, we investigate the notion of transformed measures.
First, we see that the principal pair satisfies a second-order partial differential equation.
Applying the Ito formula to \eqref{eqn:RN_X}, we have
\begin{equation*}
\frac{dL_{t}}{L_t}
=\beta\,dt+\left(\frac{1}{2}\sum_{i,j=1}^{N}a_{ij}(\partial_{ij}\phi)\phi^{-1}+\sum_{i=1}^{N}k_i(\partial_i\phi)\phi^{-1}\right)\!(X_t)\,dt
+\left(\sum_{i=1}^{N}(\partial_i\phi)\phi^{-1}\sigma_i\right)\!(X_{t})\cdot dW_{t}
\end{equation*}
and by \eqref{eqn:RN_SDE}, we know
$$\frac{dL_{t}}{L_t}=r(X_{t})\,dt+\rho_t\cdot dW_{t} \; .$$
By comparing these two equations, we obtain
\begin{equation}\label{eqn:rho}
\begin{aligned}
&\mathcal{L}\phi(x)=-\beta\phi(x)\,,\\
&\rho_{t}=\left(\sum_{i=1}^{N}(\partial_i\phi)\phi^{-1}\sigma_i\right)\!(X_{t})=\left(\sigma^*\cdot\frac{\nabla\phi}{\phi}\right)\!(X_{t})\;.
\end{aligned}
\end{equation}

\begin{theorem}
Let $(\beta,\phi)$ be
the principal pair. Then $(\beta,\phi)$
satisfies 
$\mathcal{L}\phi=-\beta\phi.$
\end{theorem} 
\noindent However, it is not true that any solution pair $(\lambda,h)$ with $h>0$ can serve as a principal pair. Refer to Theorem \ref{thm:prin_admiss} below.

Transformed measures are defined by the following way.
Let $(\lambda,h)$ be any solution pair of $\mathcal{L}h=-\lambda h$ with $h>0.$
It is easily checked that
$$e^{\lambda t-\int_{0}^{t} r(X_{s})ds}\,h(X_{t})\,h^{-1}(X_0)$$
is a local martingale under $\mathbb{Q}.$
When this is a martingale,
$(\lambda,h)$ can induce a new measure 
by setting the Radon-Nikodym derivative by $e^{\lambda t-\int_{0}^{t} r(X_{s})ds}\,h(X_{t})\,h^{-1}(X_0).$
This new measure can serve as an objective measure.
\begin{definition}
We say a solution pair $(\lambda,h)$ of $\mathcal{L}h=-\lambda p$ is an admissible pair, or
we say $h$ is an admissible function with respect to $\lambda$ 
if
$e^{\lambda t-\int_{0}^{t} r(X_{s})ds}\,h(X_{t})\,h^{-1}(X_0)$ is a martingale.
\end{definition}

\begin{theorem}\label{thm:prin_admiss}
A principal function is an admissible function. Conversely, an admissible function can serve as a principal function.
\end{theorem}

\begin{definition}
Let $(\lambda,h)$ be an admissible pair.
A measure obtained
from the risk-neutral measure $\mathbb{Q}$
by the Radon-Nikodym derivative 
$$\left.\frac{\,d\,\cdot\,}{d\mathbb{Q}}\right|_{\mathcal{F}_{t}}=e^{\lambda t-\int_{0}^{t} r(X_{s})ds}\,h(X_{t})\,h^{-1}(X_0)$$
is called
{\em the transformed measure} with respect to the admissible function $h$ or the admissible pair $(\lambda,h).$
\end{definition}
\noindent We have the following proposition by \eqref{eqn:Girsanov} and \eqref{eqn:rho}.
\begin{proposition} \label{prop:dynamics_under_P}
A process $B_{t}$ defined by
$$dB_{t}=-\left(\sigma^*\cdot\frac{\nabla h}{h}\right)\!(X_{t})\,dt+dW_{t}$$
is a Brownian motion under the transformed measure with respect to $h.$
Furthermore,
$X_{t}$ follows
\begin{equation*}
\begin{aligned}
dX_{i}(t)=\left(k_i+a_i\cdot\frac{\nabla h}{h}\right)(X_t)\,dt+\sigma_i(X_{t})\cdot dB_{t} \\
\end{aligned}
\end{equation*}
where $a_i=(a_{i1},\cdots,a_{iN}).$
\end{proposition}

Even when $e^{\lambda t-\int_{0}^{t} r(X_{s})ds}\,h(X_{t})\,h^{-1}(X_0)$ is not a martingale, we can consider the diffusion process corresponding to the operator
$$\mathcal{L}_0+a\cdot\frac{\nabla h}{h}\cdot\nabla$$
where $\mathcal{L}_0=\mathcal{L}+r.$

\begin{definition}
The diffusion process corresponding to the operator
$$\mathcal{L}_0+a\cdot\frac{\nabla h}{h}\cdot\nabla$$
is called the diffusion process {\em induced by} $h.$
\end{definition}

\section{Recurrence and transience}
\label{sec:recur_trans}
As a mathematical preliminary, we review the relationship between recurrence/transience and criticality. 
This section is indebted to \cite{Pinsky}. 
For convenience, we put $$\mathcal{G}=\mathcal{G}_\lambda=\mathcal{L}+\lambda.$$
Define 
$$C_\mathcal{G}=\left\{\,h\in C^{2}(D)\,\left|\,\mathcal{G}h=0\,,\;h(\cdot)>0 \right.\right\}\,.$$
\begin{definition} We say
\begin{itemize}[noitemsep,nolistsep]
\item[\textnormal{(i)}] $\mathcal{G}$ is {\em subcritical} if it possess a Green function,
\item[\textnormal{(ii)}] $\mathcal{G}$ is {\em critical} if it is not subcritical, but $C_\mathcal{G}$ is not empty,
\item[\textnormal{(iii)}] $\mathcal{G}$ is {\em supercritical} if it it neither critical nor subcritical.
\end{itemize}
\end{definition}

\begin{theorem} \label{thm:critical_one_dim}
If $\mathcal{G}$ is critical, then $C_\mathcal{G}$ is one-dimensional.
\end{theorem}

We are interested in a solution pair $(\lambda,h)$ of
$\mathcal{L}h=-\lambda h$ with positive function $h.$
\begin{theorem}
There exists a number $\overline{\beta}>0$ such that 
$\mathcal{L}+\lambda$ is subcritical for $\lambda<\overline{\beta},$ supercritical for $\lambda>\overline{\beta}$ and either critical or subcritical for $\lambda=\overline{\beta}.$
\end{theorem}
\noindent The assumption that $r(\cdot)$ is nonnegative is essential for this theorem. See page 148 in Pinsky for proof.
Thus, $\mathcal{L}h=-\lambda h$ always has a solution pair $(\lambda,h)$ with $h>0.$

We combine the notion of criticality with the notions of recurrence and transience.
For an open set $U$ in $\mathbb{R}^N,$ we set
$\sigma_U:=\inf\{t\geq0|X_t\notin U\}.$

\begin{definition}
The diffusion process $X_t$ on $D$ is called {\em recurrent} if $\textnormal{Prob}_x(\sigma_{B_{\epsilon}(y)}<\infty)=1$
for all $x,y\in D$ and
$\epsilon>0.$
\end{definition}

\begin{definition}
The diffusion process $X_t$ on $D$ is called {\em transient} if for all $x\in D,$
$$\textnormal{Prob}_x(X_t \textnormal{ is eventually in } D_n)=1,$$
for all $n=1,2,\cdots,$ where $\{D_n\}_{n=1}^\infty$ is a sequence of domains satisfying $D_n\subseteq D_{n+1}$ and $\overline{D_n}\subseteq D_{n+1}$ and $\cup_{n=1}^{\infty}D_n=D.$ 

\end{definition}

\noindent It is known that a diffusion process is either recurrent or transient, but cannot be both.

Criticality and subcriticality are closely related to recurrence and transience, respectively.
\begin{theorem} \label{thm:criticality_recur_trans}
If $G$ is critical, then $X_t$ is recurrent under the corresponding transformed measure.
If $G$ is subcritical, then $X_t$ is transient under the corresponding transformed measure. 
\end{theorem}

\section{The Martin integral representation}
\label{sec:Martin_repn}
As a mathematical preliminary, we review the Martin integral representation theorem.
The purpose of this section is to understand Theorem \ref{thm:Martin_repn_Markovian_kernels}.
Most contents of this section is indebted to Pinsky \cite{Pinsky}

\subsection{The Martin kernel}
Assume that $\mathcal{G}$ is subcritical and denote the Green's function by $G(x,y).$
For fixed $\xi,$ define the {\em Martin kernel} by
\begin{equation*}
k(x;y)=\left\{
\begin{aligned}
&\frac{G(x,y)}{G(\xi,y)}\;,\quad&&y\neq x\,,\;y\neq \xi\;, \\
&\quad 0\;, &&y=\xi\,,\;x\neq \xi\;,\\
&\quad 1\;,&&y=x=\xi\;.\\
\end{aligned}\right.
\end{equation*}
\noindent A sequence $\{y_n\}_{n=1}^{\infty}$ with no accumulation point in $D$ will be called a {\em Martin sequence} if $k(x;y_n)$ converges as $n$ approaches to infinity to a limit for all $x\in D.$
It is known that the limit $\lim_{n\rightarrow\infty}k(x;y_n)$ is in $C_\mathcal{G}.$
If $\{y_n\}_{n=1}^{\infty}$ and $\{y_n'\}_{n=1}^{\infty}$ are Martin sequences and 
$\lim_{n\rightarrow\infty}k(x;y_n)=\lim_{n\rightarrow\infty}k(x;y_n'),$
then two Martin sequences are called {\em equivalent}.
The collection of such equivalence classes is called the {\em Martin boundary} and will be denoted by 
$\partial.$
A point on the Martin boundary will be denoted by $\gamma.$
An element of $C_\mathcal{G}$ corresponding to $\gamma$ is denoted by $k(x;\gamma),$ that is,
$k(x;\gamma)=\lim_{n\rightarrow\infty}k(x;y_n),$ where $\{y_n\}_{n=1}^{\infty}$ is any representative of the equivalence class $\gamma.$
Occasionally, a Marin boundary point can be regarded as a curve in $D,$ denoted by $t\mapsto y(t),$
that is, $k(x;\gamma)=\lim_{t\rightarrow\infty}k(x;y(t)).$

For a bounded and connected open set $U$ in $D,$ define $\rho:(D\cup \partial)\times(D\cup \partial)\rightarrow[0,\infty)$ by
$$\rho(z_1,z_2)=\int_U\frac{|k(x;z_1)-k(x;z_2)|}{1+|k(x;z_1)-k(x;z_2)|}\,dx\;.$$

\begin{theorem}
$\rho$ is a metric on $(D\cup \partial).$ The relative topology on $D$ induced by $\rho$ is equivalent to the Euclidean topology. Under $\rho,$ $D$ is open, $\partial$ is the boundary of $D$ and $(D\cup \partial)$ is compact.
\end{theorem}
\noindent The topology on $(D\cup \partial)$ by $\rho$ is called the {\em Martin topology}.
We will use the notation `$\textnormal{Lim}$' to denote convergence in the Martin topology and to distinguish it from convergence in the Euclidean topology which is denoted by `$\lim$'.

A function in $u\in C_\mathcal{G}$ is called {\em minimal} if whenever $v\in C_\mathcal{G}$ and $v\leq u,$
then $v=cu$ for some constant $c\in(0,1].$
A point $\gamma$ in $\partial$ is called a {\em minimal Martin boundary} point if $k(x;\gamma)$ is minimal.
The collection of all minimal Martin boundary points is called the {\em minimal Martin boundary} and will be denoted by $\Lambda.$  

The following theorem is known as the {\em Martin integral representation theorem}.
\begin{theorem} 
Let $\mathcal{G}$ be subcritical.
Then for each $h\in C_\mathcal{G},$ there exists a unique finite measure $\mu_h$ on the minimal Martin boundary $\Lambda$  such that
$$h(x)=\int_{\Lambda}k(x;y)\,d\mu_h(y)\;.$$
Conversely, for each finite measure $\mu$ supported on the minimal boundary $\Lambda,$
$$h(x)\equiv\int_{\Lambda}k(x;y)\,d\mu(y)\;$$
is in $C_\mathcal{G}.$ 
\end{theorem}

\begin{corollary}
$u\in C_\mathcal{G}$ is minimal if and only if $u(x)=k(x;\gamma)$ for some $\gamma\in\Lambda.$
\end{corollary}

\subsection{Admissible boundary}
We are interested in an admissible function, 
which is a positive solution $h$ of $\mathcal{L}h=-\lambda h$ such that
$$e^{\lambda t-\int_{0}^{t} r(X_{s})ds}\,h(X_{t})\,h^{-1}(X_0)$$
is a martingale.
For this purpose, we exclude positive solutions which do not induce martingales.
Let $\tau_D$ be the exit time of the process $X_t$ from the domain $D.$
Then, 
it is a martingale if and only if
the diffusion process $X_t$ induced by $h$ does not explode, that is,
$$\text{Prob}(\tau_D=\infty)=1\;.$$
The symbol $h$ will be used for a solution of $\mathcal{L}h=-\lambda h$ with $h>0,$ and
the symbol $\phi$ will be used for an admissible function

The collection of elements $\gamma\in\Lambda$ such that 
$k(x;\gamma)$ is admissible 
is called an {\em admissible Martin minimal boundary} for $\mathcal{G}$ and will be denoted by $\Gamma.$ 
The following theorem is the main result of this article.
\begin{theorem} \label{thm:Martin_repn_Markovian_kernels}
\textnormal{(The Martin integral representation of Markovian pricing kernels)}\newline
For any admissible function $\phi,$
there exists a unique finite measure $\mu_\phi$ on the admissible Martin minimal boundary $\Gamma$ such that
$$\phi(x)=\int_{\Gamma}k(x,y)\,d\mu_\phi(y)\;.$$
Conversely, for each finite measure $\mu$ supported on $\Gamma,$
$$\phi(x)\equiv\int_{\Gamma}k(x,y)\,d\mu(y)\;$$
is an admissible function.
\end{theorem}

\noindent Any admissible function can be written as a integral of admissible minimal functions with respect to a finite measure on the admissible minimal Martin boundary.

\section{Limiting distribution}
\label{sec:lim_disri}
In this section, we demonstrate the economic or financial meaning of the measure $\mu$ on the admissible boundary $\Gamma.$  
When a process $X_t$ is positive recurrent, the meaning of the limiting distribution of $X_t$ as $t$ approaches to infinity is clear.
However, if $X_t$ is transient, then the limiting distribution does not exist in the usual sense.
While investigating limiting behavior, this phenomenon makes us harder to study transient processes.
The Martin representation is useful to overcome this. 
Indeed, the measure $\mu$ can be viewed as an `extended meaning' of limiting distribution.  

\begin{theorem}
Let $\gamma\in\Gamma$ and let $k(x;\gamma)$ be the corresponding admissible minimal function in $C_\mathcal{G}.$ 
Let $\mathbb{P}$ be the transformed measure with respect to $k(x;\gamma).$ Then 
$$\mathbb{P}\left(\underset{t\rightarrow\infty}{\textnormal{Lim}}\,X_t=\gamma\right)=1\;.$$
\end{theorem}
\noindent This implies that when the principal function of the market is an admissible minimal function, then the long-term behavior of the state variable $X_t$ is asymptotically equal to the corresponding Martin curve.

We now investigate the long-term behavior of $X_t$ for an arbitrary principal function.
We know that any principal function can be written as a integral of admissible minimal functions with respect to a finite measure $\mu$ by Theorem \ref{thm:Martin_repn_Markovian_kernels}.
In this case, the long-term distribution of the state variable under the objective measure can be expressed by the following way.

\begin{theorem}\label{thm:repn_prin}
Let $\phi$ be an principal function and let $\mu_\phi$ be the corresponding finite measure on $\Gamma.$
Denote the transformed measure with respect to $\phi$ by $\mathbb{P}.$  Then
$$\mathbb{P}\left(\underset{t\rightarrow\infty}{\textnormal{Lim}}\,X_t\in A\right)=\phi^{-1}(x)\cdot\int_{A}k(x;y)\,d\mu_\phi(y)$$
for any measurable set $A$ in $\Gamma.$
\end{theorem}

\noindent In conclusion, the measure $\mu$ is the long-term belief on $X_t$ of the market participants on the objective measure.
For more details about the limiting distribution, refer to Pinsky.

\section{Ross recovery theorem}
\label{sec:Ross_recovery}
In this section, we investigate the Ross recovery theorem in the view of the Martin representation.
The purpose of the Ross recovery theorem is to find the objective measure under Assumption 1, 2 and 3 with the assumption that the $Q$-dynamics of $X_t$ and the interest rate function $r(\cdot)$ are known ex ante.

\subsection{Multi-dimensional state variable}
We first explore the Ross recovery theorem with a multi-dimensional state variable.
By Theorem \ref{thm:critical_one_dim} and \ref{thm:criticality_recur_trans}, we have the following proposition.

\begin{proposition}
If existent, there is a unique admissible pair $(\beta,\phi)$ such that $X_t$ is recurrent under the transformed measure with respect to the pair $(\beta,\phi).$
\end{proposition}

\begin{theorem}
If $X_t$ is recurrent under the objective measure, then recovery is possible.
\end{theorem}

When studying Ross recovery with the multi-dimensional state variable $X_t,$ 
transient recovery is inevitable for the following two reasons.
First, if at least one component of $X_t$ is transient, then the process $X_t$ is transient.
In practical financial markets, at least one component of the state variable is usually transient.
For example, consider a state variable $X_t=(X_1(t),X_2(t),X_3(t))$ where $X_1(t)$ is S\&P 500 index, $X_2(t)$ is the volatility of $X_1(t)$ and $X_3(t)$ is the interest rate process, in which case, $X_1(t)$ is transient under the objective measure.
This observation provides a state variable example to study transient recovery.

Second, even though $X_t$ is componentwise recurrent, $X_t$ can be transient.
More precisely, even though each component $X_i(t)$ is recurrent for all $i\in\{1,2,\cdots,N\},$ the process $X_t$ itself can be transient. As is well known, a $N$-dimensional Brownian motion $B_t=(B_1(t),B_2(t),\cdots,B_N(t))$ is transient for $N\geq3$ even though $X_t$ is componentwise recurrent. 
Consider the state variable $X_t=(X_1(t),X_2(t),X_3(t))$ 
where $X_1(t)$ is the volatility of a stock, $X_2(t)$ is the volatility of another stock
and $X_3(t)$ is the interest rate.
We want to find an objective measure such that each $X_t$ is componentwise recurrent because the interest rate and the volatilities are recurrent in the actual market. 
This point provides an implication that the study of transient recovery is inevitable
when the state variable is multidimensional.

There are only a few studies on the Ross recovery theorem for the multi-dimensional transient state variable $X_t.$
We believe that this is because multi-dimensional transient recovery is more difficult to research.
In this sense, the Martin representation can be a bright idea.

\begin{theorem}
If the state variable is transient under the objective measure, then Ross recovery is equivalent to choosing a number $\beta$ with $\beta\leq\overline{\beta}$ and a finite measure $\mu$ on $\Gamma_\beta.$
In this case, the long-term belief on $X_t$ of the market participants is expressed by
$$\mathbb{P}\left(\underset{t\rightarrow\infty}{\textnormal{Lim}}\,X_t\in A\right)=\phi^{-1}(x)\cdot\int_{A}k(x;y)\,d\mu_\phi(y)$$
for any measurable set $A$ in $\Gamma.$
\end{theorem}
\noindent Any information that determines an objective measure contains the principal factor $\beta$ and the finite measure $\mu.$ Conversely, a number and a finite measure on the admissible Martin minimal boundary determine the objective measure.
This theorem implies that when the state variable $X_t$ is transient under the objective measure, we need to know the value $\beta$ and the `limiting distribution' of $X_t$ in the Martin topology under the objective measure.
It seems awkward that one would know the `limiting distribution' of $X_t$ in the Martin topology under the objective measure, but occasionally one would know the `limiting distribution'.
To see this, we investigate the following examples.

\begin{example} \textnormal{(Multidimensional Brownian motion)}

We investigate how the Martin representation is applied to Ross recovery when the state variable is a (scaled) multidimensional Brownian motion.
See \cite{Pinchover} for more details.
$$X_t=\sqrt{2}\,W_t=\sqrt{2}\,(W_1(t),\cdots,W_N(t))\;.$$
Assume that the interest rate is a constant and that the value $\beta$ is known. Let $\lambda:=-r+\beta.$
The corresponding second-order equation is 
$$\mathcal{G}h=\Delta h+\lambda h=0\;.$$

If $\lambda=0$ and $N\geq3,$ then 
the Martin boundary consists of only one point.
If $\lambda<0,$ then 
the $S^{N-1}$ is the Martin boundary, which is the sphere at infinity. 
For any $\gamma\in S^{N-1}=\{z\in\mathbb{R}^N:|z|=1\},$
$$k(x;\gamma)=e^{\sqrt{-\lambda}\,\gamma\cdot x}$$ is the corresponding minimal function and
a corresponding Martin curve is $t\mapsto \gamma t\,.$  
It can be shown that every Martin boundary is an admissible minimal boundary.
Therefore, by Theorem \ref{thm:Martin_repn_Markovian_kernels},
$\phi$ is admissible if and only if there is a finite measure $\mu$ on $S^{N-1}$ such that
$$\phi(x)=\int_{S^{N-1}}e^{\sqrt{-\lambda}\,\gamma\cdot x}\,d\mu(\gamma)\;.$$

As a simple application to Ross recovery, suppose that  $N=2,$ the interest rate is $r=1$ and 
 $\lambda=-1+\beta<0.$ 
Let $X_1(t)$ be the log of a stock price and let $X_2(t)$ be the log of a foreign exchange rate.  
Under the risk-neutral measure, suppose that $X_1(t)=\sqrt{2}\,W_1(t)$ and $X_2=\sqrt{2}\,W_2(t).$
It is a reasonable belief that under the objective measure, $X_1(t)$ goes to infinity as $t$ goes to infinity and $X_2(t)$ is recurrent.   
This implies that the limiting distribution of $X_t$ in the Martin topology under the objective measure is 
$$\mathbb{P}\left(\underset{t\rightarrow\infty}{\textnormal{Lim}}\,(X_1(t),X_2(t))=\gamma\right)=1$$
where $\gamma$ induces a Martin curve $t\mapsto (t,0).$
Clearly, this $\mathbb{P}$ is the only transformed measure satisfying the reasonable belief.
The corresponding minimal function is
$\phi(x)=e^{\sqrt{-\lambda}\,x_1}$ where $x=(x_1,x_2).$
$B_t:=W_t-(\sqrt{-\lambda}\,t,0)$ is a Brownian motion under 
the transformed measure with respect to this $\phi.$
Thus, we obtain that under the objective measure, $X_t$ follows
$$X_t=\sqrt{2}\,W_t=\sqrt{2}\,B_t+(\sqrt{-2\lambda}\,t,0)\;.$$

For another example, suppose that  $N=2,$ the interest rate is $r=1$ and 
 $\lambda=-1+\beta<0.$ 
Let $X_1(t)$ be the log of a stock price and let $X_2(t)$ be the log of another stock price.  
Under the risk-neutral measure, suppose that $X_1(t)=\sqrt{2}\,W_1(t)$ and $X_2=\sqrt{2}\,W_2(t).$
It is a reasonable guessing that under the objective measure, both $X_1(t)$ and $X_2(t)$ go to infinity as $t$ approaches infinity. 
There is an infinitely number of admissible functions satisfying this condition, thus we need more information for recovery. Assume that the long-term ratio $$\frac{p}{q}=\lim_{t\rightarrow\infty}\frac{X_1(t)}{X_2(t)}$$ is known with $p,q>0$ and $p^2+q^2=1.$
This implies that the limiting distribution of $X_t$ in the Martin topology under the objective measure is 
$$\mathbb{P}\left(\underset{t\rightarrow\infty}{\textnormal{Lim}}\,(X_1(t),X_2(t))=\gamma\right)=1$$
where $\gamma$ induces a Martin curve $t\mapsto (pt,qt).$
Clearly, this $\mathbb{P}$ is the only transformed measure satisfying the condition.
The corresponding minimal function is
$\phi(x)=e^{\sqrt{-\lambda}\,(px_1+qx_2)}$ where $x=(x_1,x_2).$
Therefore, under the objective measure, $X_t$ follows
$$X_t=\sqrt{2}\,W_t=\sqrt{2}\,B_t+(\sqrt{-2\lambda}\,pt,\sqrt{-2\lambda}\,qt)\;.$$
\end{example}

\begin{example} \textnormal{(State variable with constant coefficients)}

Any elliptic operator with constant coefficients may be reduced to the form $\Delta+\lambda,$ where $\lambda$ is a constant, by suitable transformations which preserve the Martin boundary. 
Thus, the Martin boundary is either the sphere $S^{N-1}$ or one point.
By the same argument above, there exists a unique admissible function such that the $X_t$ is componentwise recurrent under the corresponding transformed measure. In this case, the Martin boundary is one point.

There exists a unique admissible function such that the $X_1(t)$ goes to infinity as $t$ approaches to infinity and $X_2(t),\cdots,X_N(t)$ are (componentwise) recurrent
under the corresponding transformed measure when the value $\beta$ is known. In this case, $t\mapsto (t,0,\cdots,0)$ is a corresponding Martin curve.

There exists a unique admissible function such that the $X_1(t),\cdots,X_k(t)$ goes to infinity as $t$ approaches to infinity with the long-term ratio
$$\frac{p_{i}}{p_{i+1}}=\lim_{t\rightarrow\infty}\frac{X_i(t)}{X_{i+1}(t)}\quad\text{ for } i=1,2,\cdots,k-1$$
and $X_{k+1}(t),\cdots,X_N(t)$ are (componentwise) recurrent
under the corresponding transformed measure when the value $\beta$ is known. In this case, $t\mapsto (p_1t,\cdots,p_kt,0,\cdots,0)$ is a corresponding Martin curve when the value $\beta$ is known.
\end{example}

\begin{example} \textnormal{(Two-dimensional OU processes)}

This example is indebted to \cite{Cranston}.
Consider a two-dimensional Ornstein-Uhlenbeck process $X_t.$
$$dX_t=dW_t+BX_t\,dt\;,$$
where $B$ is a $2\times2$ nonsingular matrix. 
We assume that interest rate is a constant and is equal to the value $\beta.$ 
The corresponding second-order equation is
$$\mathcal{G}h(x)=\frac{1}{2}\Delta h(x)+<Bx,\nabla h(x)>=0\;.$$
Despite the fact that $X_t$ is a Gaussian process, a direct calculation of the Martin boundary does
not appear easy.
If both eigenvalues of $B$ have non-positive real part, $X_t$ is recurrent. The generator $\mathcal{G}$ is critical.

If both eigenvalues of $B$ have positive real part, then 
the minimal Martin boundary is $S^1$ and the minimal Martin curves are of the form
$$t\mapsto e^{Bt}\cdot\gamma\quad\textnormal{ for } \gamma\in S^1\;.$$ 
Let $C_B:=\int_0^\infty e^{-Bs}e^{-B^*s}\,ds$ and denote two eigenvalues of $B$ by $z_1$ and $z_2.$ 
Define
$$K_B(x,t;\gamma):
=e^{(z_1+z_2)t}\exp\left(-\frac{1}{2}(e^{Bt}\gamma-x)^{*}C_B^{-1}(e^{Bt}\gamma-x)
-\gamma^*C^{-1}\gamma\right)\;.$$
Then, the minimal function corresponding to $\gamma\in S^1$ is 
$$k_B(x;\gamma)=c_\gamma^{-1}\int_{-\infty}^{\infty}K_B(x,s;\gamma)\,ds\;,$$
where
$c_\gamma=\int_{-\infty}^{\infty}K_B(0,s;\gamma)\,ds.$

When $B$ has one positive and one negative eigenvalue, the situation is more complicated.
Let $z_2<0<z_1.$
By changing an orthogonal change of coordinates, $B$ can be put in subtriangular from, thus we may assume 
\begin{equation*}
B=\begin{bmatrix}
    z_2  &  0  \\
    b     & z_1 \\
\end{bmatrix}\;.
\end{equation*}
Define 
\begin{equation*}
\hat{B}=\begin{bmatrix}
    -z_2  &  0  \\
    b     & z_1 \\
\end{bmatrix}\;,
\end{equation*}
then it can be shown that 
the minimal Martin boundary is $S^1$ and the minimal Martin curves are of the form
$$t\mapsto e^{\hat{B}t}\cdot\gamma\quad\textnormal{ for } \gamma\in S^1\;.$$ 
The minimal function corresponding to $\gamma\in S^1$ is 
$$k_{\hat{B}}(x;\gamma)=\hat{c}_\gamma^{-1}\int_{-\infty}^{\infty}K_{\hat{B}}(x,s;\gamma)\,ds\;,$$
where
$\hat{c}_\gamma=\int_{-\infty}^{\infty}K_{\hat{B}}(0,s;\gamma)\,ds.$
In conclusion, the Martin boundary is either the sphere $S^{N-1}$ or one point.

Let $\mathcal{G}$ be critical.
By the same argument, there exists a unique admissible function such that the $X_1(t)$ goes to infinity as $t$ approaches to infinity and $X_2(t),\cdots,X_N(t)$ are (componentwise) recurrent
under the corresponding transformed measure when the value $\beta$ is known. In this case, $t\mapsto (t,0,\cdots,0)$ is a corresponding Martin curve.

There exists a unique admissible function such that the $X_1(t),\cdots,X_k(t)$ goes to infinity as $t$ approaches to infinity with the long-term ratio
$$\frac{p_{i}}{p_{i+1}}=\lim_{t\rightarrow\infty}\frac{X_i(t)}{X_{i+1}(t)}\quad\text{ for } i=1,2,\cdots,k-1$$
and $X_{k+1}(t),\cdots,X_N(t)$ are (componentwise) recurrent
under the corresponding transformed measure when the value $\beta$ is known. In this case, $t\mapsto (p_1t,\cdots,p_kt,0,\cdots,0)$ is a corresponding Martin curve when the value $\beta$ is known.
\end{example}

From these observations, we obtain the following theorem.
\begin{theorem}
Let $\mathcal{G}$ be subcritical. Assume that the Martin boundary is $S^{N-1}.$
If $X_1(t)$ goes to infinity and $X_2(t),\cdots,X_N(t)$ are recurrent under the objective measure, then recovery is possible.
\end{theorem}

We now shift our attention to the choice of the principal factor $\beta.$
When $X_{t}$ is transient, to recover the objective measure, we confront a problem of determining the value $\beta.$ 
How can we choose the value?
One way is to use the long-term yield of bonds, which is defined by 
\begin{equation*}
\lim_{t\rightarrow \infty}\left(-\frac{1}{t}\cdot\log \mathbb{E}^{\mathbb{Q}}\left[
e^{-\int_{0}^{t}r(X_{s})\,ds}\right]\right)\;.
\end{equation*}
See \cite{Martin} and \cite{Qin14b} as a reference.

We may use empirical data for $\beta.$ 
Finding $\beta$ is an important topic in finance and economics. There is
a vast amount of literature on the study of  and the theoretical and practical
methods of finding ;  Bansal and Yaron \cite{Bansal04}, Breeden \cite{Breeden79} and Campbell
and Cochrane \cite{Campbell99}. By these methods, we can obtain proper empirical data
of $\beta.$

\subsection{One-dimensional state variable}
We explore the Ross recovery theorem for one-dimensional state variable $X_t.$
In this case, the Martin boundary is very simple, thus Ross recovery can be easily analyzed.
We can singles out a unique recovery out of all feasible recoveries which also has economic meaning (approaching to one boundary).
For an elementary approach to one dimensional case without the Martin theory, refer to \cite{Park14b}.

\begin{theorem}
Let $N=1$ and let $\mathcal{G}=\mathcal{L}+\beta$ be subcritical.
Then the Martin minimal boundary is $\Lambda=\{-1,1\}.$ 
\end{theorem}
\noindent 
Denote the left and right boundary of the range of $X_t$ by $a$ and $b,$ respectively. 
The diffusion process induced by the minimal function $k(x;1)$
is transient and goes to the right boundary $b.$
Clearly, $k(x;1)$ is the only function in $C_\mathcal{G}$ which induce the diffusion process going to the right boundary as $t$ goes to infinity. 
Similarly, the diffusion process induced by the minimal function $k(x;-1)$
is transient and goes to the left boundary $a.$

\begin{theorem}
\textnormal{(Transient recovery)}
Suppose we know the value $\beta$ and let $\mathcal{G}=\mathcal{L}+\beta$ be subcritical.
If only one of $k(x;1)$ and  $k(x;-1)$ is admissible, then
we can recover the objective measure from the risk-neutral measure.
\end{theorem}

We investigate another way for transient recovery. 
When the principal factor $\beta$ is known, if it exists, there is a unique 
admissible function such that $X_{t}$ goes to the right (or left) boundary as $t$ approaches to infinity under the corresponding transformed measure.
The corresponding Martin boundary point is $\{1\}$ (or $\{-1\}$).

\begin{theorem}\label{thm:transient_recovery}\textnormal{(Transient recovery)}
Suppose we know the value $\beta.$
If $X_{t}$ goes to the right boundary (or left boundary) as $t$ approaches to infinity under the 
objective measure $\mathbb{P},$ then we can recover the objective measure  
from the risk-neutral measure.
\end{theorem}

\noindent 
When the state variable is a stock price, this theorem is useful.
The left and right boundary of a stock price is $0$ and $\infty,$ respectively (for example, the Black-Scholes model).
The stock price process usually goes to infinity under the objective measure, thus we can recover the objective measure when the value $\beta$ is known. 

If $X_{t}$ goes to both boundaries under the objective measure, we 
need more information for recovery. 
Let $p,q$ be two positive numbers with $p+q=1.$
Then, if existent, there is a unique admissible function such that
$X_t$ goes to the left boundary with probability $p$ and 
$X_t$ goes to the right boundary with probability $q.$
The corresponding measure $\mu$ on the Martin boundary $\{-1,1\}$ satisfies $\mu(-1)=p$ and $\mu(1)=q.$

\section{Long-time behavior of cash flows}
\label{sec:long-term}
Borovicka, Hansen and Scheinkman.
investigated the long-term decay (or growth) rate of cash flows in \cite{Borovicka14}.
In this section, we investigate their work in the view of the Martin representation.

\begin{proposition}
Assume that $\mathcal{G}$ is subcritical and let $(D\cup \partial)$ be the corresponding Martin space. 
Let $\phi$ be an admissible function and denote the corresponding measure by $\mu_\phi.$
If $g$ is a continuous function on $(D\cup \partial),$ then 
$$\lim_{t\rightarrow\infty}\mathbb{E}^{\mathbb{P}}[g(X_t)]=\frac{1}{\phi(x)}\int_{\Gamma}g(y)k(x;y)\mu_\phi(y)$$
where $\mathbb{P}$ is the transformed measure with respect to $\phi.$
\end{proposition}
\noindent It is clear if $g=\chi_B$ for any measurable set $B$ in $(D\cup \partial)$ by Theorem \ref{thm:repn_prin}. The proof is completed by the density argument.

Let $\phi$ be an admissible function of $\mathcal{G}\phi=\mathcal{L}\phi+\beta\phi=0$ and let $\mathbb{P}$ be the corresponding transformed measure. Then the pricing operator in \eqref{eqn:pricing_oper} becomes
$$P_t(f)(x):=\mathbb{E}^{\mathbb{Q}}_x\left[e^{-\int_0^tr(X_s)\,ds}f(X_t)\right]
=\phi(x)e^{-\beta t}\cdot\mathbb{E}^{\mathbb{P}}[(\phi^{-1}f)(X_t)]\;.$$
By the previous proposition, we have the following theorem.
\begin{theorem}
If $(\phi^{-1}f)(\gamma):=\underset{x\rightarrow\gamma}{\textnormal{Lim}}(\phi^{-1}f)(x)$ exists for all $\gamma\in\partial,$ then  
$$\lim_{t\rightarrow\infty}e^{\beta t}\,p_t=\int_{\Gamma}(\phi^{-1}f)(y)\,k(x;y)\mu_\phi(y)$$
where $p_t=P_t(f)(x).$
\end{theorem}
\noindent Thus, if one can find an admissible pair $(\beta,\phi)$ such that 
$(\phi^{-1}f)(\gamma)$ exists for all $\gamma\in\partial$
and is not identically equal to zero on the support of $\mu_\phi,$
then the long-term decay (or growth) exponential rate is obtained.

\section*{Acknowledgments}

The author is supported by the MacCracken Fellowship at the Courant Institute, New York University.

\section{Conclusion}
We described all the possible beliefs of market participants on objective measures under Markovian environments
when a risk-neutral measure is given.
To achieve this, we employed the Martin integral representation of Markovian pricing kernels and then offered economic and financial implications of the representation. 
As applications, the Ross recovery theorem and the long-term behavior of cash flows were discussed.

We investigated 
the Ross recovery with a multi-dimensional Makovian diffusion $X_t$ that is transient under the objective measure.
In addition, as a special case, one-dimensional state variable was analyzed. 
Ross recovery is equivalent to choosing a principal factor $\beta$ and a finite measure $\mu$ on $\Gamma_\beta$
where $\Gamma_\beta$ is the corresponding Marin admissible boundary.
As an another application, we investigated long-term behavior of cash flows
in the view of the Martin representation.

This article offers a theoretical approach for multi-dimensional transient Ross recovery.
Usually, it is very difficult to obtain the precise Martin boundary for a multi-dimensional case.
For future research, it would be interesting to find the precise Martin boundary for a specific multi-dimensional model.

\end{document}